# Exact analytic expressions for electromagnetic propagation and optical nonlinear generation in finite one-dimensional periodic multilayers

Matteo Cherchi[*]

Pirelli Labs – Optical Innovation, viale Sarca 222, 20126 Milan, Italy

Translation Matrix Formalism has been used to find an exact analytic solution for linear light propagation in a finite one-dimensional (1D) periodic stratified structure. This modal approach allows to derive a closed formula for the electric field in every point of the structure, by simply imposing a convenient form for the boundary conditions. We show how to apply this result to Second Harmonic Generation (SHG) in the undepleted pump regime.

42.70.Qs, 42.79.Dj, 42.79.Nv, 42.25.Bs

## I. INTRODUCTION

It is a very long time since multilayer periodic structures have been proposed as distributed cavities for many different purposes[1,2]. They allow the engineering of field enhancement, group velocities, coherence lengths and so on. In literature their treatment is usually performed numerically, for example with the Transfer Matrix approach[3,4,5] or by a Runge-Kutta numerical integration of the Maxwell equations[6], and semi analytically through convenient approximations, like the coupled mode theory[7][8][9], the envelope function approach[10][11][12][13], the Fourier expansion of Bloch waves[14][15][16] or the recently proposed quasinormal mode expansion[17].

---

[*] cherchi@cantab.net

In this paper we will derive exact analytical expressions both for the transmission and for the local field distribution of finite periodic structures using the modal approach as proposed by Russell et al.[14]. This approach greatly simplifies the heavy numerical calculation needed in other schemes. Physically a purely 1D system represents, in principle, a scalar problem with 2 degrees of freedom: namely going from left to right and going from right to left. We will show how, working always in the basis of the eigenmodes, in each self similar section 2 complex numbers completely determine an exact analytical solution for the propagation. Self similarity means translation invariance, i.e. that the structure can be seen as a succession of replicas of the same unit cell. An uniform medium clearly features a continuous translational invariance (i.e. any point it is equivalent to any other point) and its eigenmodes are, of course, the forward and backward propagating plane waves. On the other hand a periodic medium features a discrete translation invariance (i.e. points a period apart are equivalent to each other) and its eigenmodes are well known to be the progressive and regressive Bloch waves.

By definition, these eigenmodes propagate unperturbed in their proper medium, reducing the problem to a simple analysis of the boundary conditions, coming from the world of plane waves, passing through the world of Bloch waves to couple back to the realm of plane waves.

The same approach can be easily extended to the nonlinear regime whenever the linear propagation of the input fields is not significantly affected by the nonlinear interactions. As an example we will study the case of SHG in the undepleted pump regime.

## II. BLOCH MODES

Let's consider a structure which period is composed of *M* different layers with refractive index $n_j$ and thickness $d_j$ ($j = 1…M$). The stack period will be $\Lambda \equiv \sum_j d_j$ (see Fig. 1). For simplicity we will focus on the purely 1D problem of propagation normal to the layer (anyway it is straightforward to extend our analysis to oblique incidence[14]).

It is a well established result that, in an uniform medium, the general solution for the scalar wave equation can be written as a superposition of plane waves. Following Russell's Translation Matrix Formalism[14] we can choose a particular form for this linear combination and write the field in the $j^{th}$ layer of the $m^{th}$ period as a superposition of a sine and a cosine centered in the middle of the layer:

$$E_{j,m}(z) = a_{j,m} \cos[\boldsymbol{b}_j(z - z_{j,m})] + b_{j,m} \sin[\boldsymbol{b}_j(z - z_{j,m})]/\boldsymbol{q}_j , \qquad (1)$$

where $a_{j,m}$ and $b_{j,m}$ are complex constants, $\boldsymbol{b}_j \equiv 2\boldsymbol{p}n_j/\boldsymbol{l}$, is the propagation constant of a plane wave in the considered layer, $\boldsymbol{q}_j \equiv \boldsymbol{b}_j \Lambda$ and

$$z_{j,m} = m\Lambda + \tfrac{1}{2}(d_j - d_1) + \sum_{i=1}^{j-1} d_i$$

is the coordinate of the center of the same layer. This particular form has been carefully chosen for algebraic convenience. The basis $\{\cos[\mathbf{b}_j(z - z_{j,m})], \sin[\mathbf{b}_j(z - z_{j,m})]/\mathbf{q}_j\}$ is specific for each single layer, is always real even for imaginary $\mathbf{b}_j$, well behaved as $\mathbf{b}_j^2$ changes sign and retains two degrees of freedom even for $\mathbf{b}_j = 0$ (actually these features are really important only in the case of oblique incidence). Also propagation within a stop band is specified by real values of $a_{j,m}$ and $b_{j,m}$. In general these constants entirely specify the field in any chosen layer.

Introducing the vectorial representation of the field states in the $j^{th}$ layer of the $m^{th}$ period:

$$\mathbf{F}_{j,m} \equiv \begin{pmatrix} a_{j,m} \\ b_{j,m} \end{pmatrix},$$

the translation matrices

$$\mathbf{M}_{ki} \equiv \begin{pmatrix} A_{ki} & B_{ki} \\ C_{ki} & D_{ki} \end{pmatrix},$$

so that $\mathbf{F}_{k,m} = \mathbf{M}_{ki}\mathbf{F}_{i,m}$, propagate the field from the layer $i$ to the adjacent layer $k \equiv \mathrm{mod}(i, M) + 1$ (where 'mod($x$, $y$)' means the remainder on division of $x$ by $y$). Their elements are:

$$\begin{aligned} A_{ki} &= c_k c_i - (\mathbf{q}_i/\mathbf{q}_k)s_i s_k; & B_{ki} &= c_i s_k/\mathbf{q}_k + c_k s_i/\mathbf{q}_i \\ D_{ki} &= c_i c_k - (\mathbf{q}_k/\mathbf{q}_i)s_i s_k; & C_{ki} &= -(\mathbf{q}_k c_i s_k + \mathbf{q}_i c_k s_i) \end{aligned},$$

having defined $c_j \equiv \cos(\mathbf{b}_j d_j/2)$ and $s_j \equiv \sin(\mathbf{b}_j d_j/2)$. These matrices are entirely real whenever $\mathbf{q}_j$ are real, which is true in the case of normal incidence.

The matrix

$$\mathbf{M} \equiv \mathbf{M}_{1M} \cdots \mathbf{M}_{32}\mathbf{M}_{21} \equiv \begin{pmatrix} A & B \\ C & D \end{pmatrix},$$

so that $\mathbf{F}_{1,m+1} = \mathbf{M}\mathbf{F}_{1,m}$, links the field of corresponding layers in adjacent periods.

By definition the Bloch modes are the eigenmodes of a periodic structure. Since the matrix $\mathbf{M}$ propagates the field from the first layer of a given period to the first layer of the next period, the vectorial representation of the Bloch mode in the first layer of the first period will be given (up to an overall phase) by the eigenstates of $\mathbf{M}$:

$$\mathbf{F}_1^\pm \equiv \frac{2}{R}\begin{pmatrix} 2B \\ D - A \pm \sqrt{(D-A)^2 + 4BC} \end{pmatrix} \equiv \begin{pmatrix} a_1^\pm \\ b_1^\pm \end{pmatrix}, \qquad (2)$$

where $R \equiv \sqrt{4B^2 + (D - A \pm \sqrt{(D-A)^2 + 4BC})^2}/q_1^2$ is a convenient normalization. We can then propagate the field from the layer $i$ to the adjacent layer $k \equiv \mathrm{mod}(i,M)+1$ to find the representation of the Bloch mode in the other layers as

$$\mathbf{F}_k^\pm \equiv \mathbf{M}_{ki}\mathbf{F}_i^\pm \equiv \begin{pmatrix} a_k^\pm \\ b_k^\pm \end{pmatrix}.$$

They correspond to the eigenvalues

$$l_\pm = \tfrac{1}{2}(D + A \pm \sqrt{(D-A)^2 + 4BC}) \equiv \exp(\pm i\mathbf{q}) \equiv \exp(\pm i\mathbf{b}\Lambda)$$

where $\mathbf{b} \equiv \arccos[\tfrac{1}{2}(D+A)]/\Lambda$ is the Bloch propagation constant. So, up to an overall phase, the eigenstates in the $j^{\text{th}}$ layer of the $m^{\text{th}}$ period can be written as

$$\mathbf{F}_{j,m}^{\pm} = \exp(\pm im\mathbf{q})\mathbf{F}_{j}^{\pm} = \exp(\pm im\mathbf{q})\mathbf{F}_{j,0}^{\pm}$$

or, in terms of the field eigenmodes:

$$E_{j,m}^{\pm}(z) = \exp(\pm im\mathbf{q})\{a_j^{\pm}\cos[\mathbf{b}_j(z-z_{j,m})] + b_j^{\pm}\sin[\mathbf{b}_j(z-z_{j,m})]/\mathbf{q}_j\} = \exp(\pm im\mathbf{q})E_{j,0}^{\pm}(z),$$

which reads that, for a Bloch mode, corresponding points of different periods feature the same field, up to a factor due to the Bloch propagation constant (that is a phase factor in the pass band and an amplitude factor in the band gap).

Let's now introduce the square waves

$$S_j(z) \equiv \text{pos}\left[d_j - \text{mod}\left(z + \frac{d_1}{2} - \sum_{i=1}^{j-1}d_i, \Lambda\right)\right],$$

where we have defined

$$\text{pos}(x) \equiv \begin{cases} 1 & x \geq 0 \\ 0 & x < 0 \end{cases}.$$

They are defined so that $S_j(z)$ is 1 in the regions with refractive index $n_j$ and zero elsewhere.

Let's define also the 'period number' function ('int' is for integer part)

$$q(z) \equiv \text{int}\left(\frac{z + d_1/2}{\Lambda}\right)$$

and the relative positions

$$\boldsymbol{d}_j(z) \equiv z - [q(z)\Lambda + z_{j,0}] = z - \left[q(z)\Lambda + \tfrac{1}{2}(d_j - d_1) + \sum_{i=1}^{j-1} d_i\right].$$

With these elements the Bloch modes can be rewritten in closed form as

$$E^{\pm}(z) = \sum_{j=1}^{M} S_j(z) E_j^{\pm}(z), \tag{3}$$

where

$$E_j^{\pm}(z) \equiv \exp[\pm i\boldsymbol{q}q(z)]\{a_j^{\pm} \cos[\boldsymbol{b}_j \boldsymbol{d}_j(z)] + b_j^{\pm} \sin[\boldsymbol{b}_j \boldsymbol{d}_j(z)]/\boldsymbol{q}_j\}.$$

It can be also convenient to represent the generic field Eq. (1) as a superposition of local plane waves (i.e. the eigenmodes of each single layer) through the vectors

$$\mathbf{Y}_{j,m} \equiv \begin{pmatrix} g_{j,m} \\ h_{j,m} \end{pmatrix} \equiv \frac{1}{2}\begin{pmatrix} 1 & i/\boldsymbol{q}_j \\ 1 & -i/\boldsymbol{q}_j \end{pmatrix}\begin{pmatrix} a_{j,m} \\ b_{j,m} \end{pmatrix} \equiv \mathbf{T}_j \mathbf{F}_{j,m}, \tag{4}$$

having rewritten

$$E_{j,m}(z) = g_{j,m} \exp[-i\boldsymbol{b}_j \boldsymbol{d}_j(z)] + h_{j,m} \exp[i\boldsymbol{b}_j \boldsymbol{d}_j(z)]. \tag{5}$$

This allows to cast Eq. (3) in the standard Bloch form $E^{\pm}(z) \equiv u_{\pm}(z)\exp(\pm i\boldsymbol{b}z)$ by defining

$$u_{\pm}(z) \equiv \sum_{j=1}^{M} S_j(z) u_j^{\pm}(z),$$

where

$$u_j^\pm(z) \equiv \exp(\mp i\mathbf{b}\, z_{j,0})\{g_j^\pm \exp[-i(\mathbf{b}_j \pm \mathbf{b})\mathbf{d}_j(z)] + h_j^\pm \exp[i(\mathbf{b}_j \pm \mathbf{b})\mathbf{d}_j(z)]\},$$

with the obvious definition of $g_j^\pm$ and $h_j^\pm$ through $\mathbf{Y}_j^\pm \equiv \mathbf{T}_j \mathbf{F}_j^\pm$.

From the definitions of $S_j(z)$ and $\mathbf{d}_j(z)$, it is clear that $u_\pm(z)$ are $\Lambda$-periodic functions.

## III. BOUNDARY CONDITIONS

### A. Perfectly periodic case

Consider now a finite structure of $N$ periods (Fig. 2) embedded, both on the left and on the right, in the medium 1, and excited on the left by a plane wave with wave vector normal to the layers, unity amplitude and null phase at $z = 0$.

Assuming that all materials are lossless, the layered structure will act as a mirror with unitary scattering matrix, reflecting a plane wave with amplitude $r$ and a phase $\mathbf{j}_r$ and transmitting a plane wave with amplitude $t$ and phase $\mathbf{j}_t$, all to be determined.

In case of nondegeneracy, the eigenmodes of the periodic structure represent a complete basis in every layer of every period (notice that at the band edge points, where $(D-A)^2 + 4BC = 0$, the eigenmodes are degenerate and the eigenstates can be a complete basis if and only if $B = C = 0$ and $D = A$). So we can write the field in every point of the periodic section as a linear combination:

$$E(z) = X_+ E^+(z) + X_- E^-(z), \tag{6}$$

where $X_\pm$ are the expansion coefficients, i.e., two complex constants to be determined. Since the Bloch waves can be always projected on the complete basis of the eigenmodes of the single layer (Eq. (5)), we can highlight the forward and backward propagating plane wave terms rewriting

$$E(z) \equiv \sum_{j=1}^{M} S_j(z)\{A_j^f(z)\exp[-i\mathbf{b}_j\mathbf{d}_j(z)] + A_j^b(z)\exp[i\mathbf{b}_j\mathbf{d}_j(z)]\},$$

where we have defined the z-dependent expansion coefficients

$$\begin{aligned}A_j^f(z) &= \{X_+ g_j^+ \exp[i\mathbf{q}q(z)] + X_- g_j^- \exp[-i\mathbf{q}q(z)]\}\exp[-i\mathbf{b}_j\mathbf{d}_j(z)] \\ A_j^b(z) &= \{X_+ h_j^+ \exp[i\mathbf{q}q(z)] + X_- h_j^- \exp[-i\mathbf{q}q(z)]\}\exp[i\mathbf{b}_j\mathbf{d}_j(z)]\end{aligned}. \qquad (7).$$

This allows to easily impose the boundary conditions in terms of the incoming and outgoing plane waves:

$$\begin{cases} A_1^f(0) = 1 \\ A_1^b(0) = r\exp(i\mathbf{j}_r) \equiv \mathbf{r}(N) \\ A_1^f(N\Lambda) = t\exp(i\mathbf{j}_t) \equiv \mathbf{t}(N) \\ A_1^b(N\Lambda) = 0 \end{cases}$$

giving:

$$\begin{cases} X_+ = \Omega_{11}^N \exp(-iN\mathbf{q})h_1^- \\ X_- = -\Omega_{11}^N \exp(iN\mathbf{q})h_1^+ \\ \mathbf{t}(N) = \Omega_{11}^N (g_1^+ h_1^- - g_1^- h_1^+) \\ \mathbf{r}(N) = -2i\Omega_{11}^N \sin(N\mathbf{q})h_1^+ h_1^- \end{cases}, \qquad (8)$$

having defined

$$\Omega_{jk}^{N} \equiv [h_j^- g_k^+ \exp(-iN\boldsymbol{q}) - h_j^+ g_k^- \exp(iN\boldsymbol{q})]^{-1}. \tag{9}$$

In Eq. (8) $\boldsymbol{t}(N)$ and $\boldsymbol{r}(N)$ are analytical expressions for the transmission and reflection of a *N*-period multilayer (analogous to those found in Ref. 2), while $X_+$ and $X_-$, once substituted in Eq. (6), give analytical expressions for the field inside the multilayer. Notice that, at the band edges, $\Omega_{11}^{N}$ diverges and, while $\boldsymbol{t}(N)$ and $\boldsymbol{r}(N)$ are well behaved functions, $X_+$ and $X_-$ diverge.

To our knowledge this is the first time that an exact analytical expression for the field in a finite multilayer structure is calculated explicitly. We believe this is due to two main reasons. First, in literature it is still possible to find papers claiming that Bloch analysis can be applied to infinite periodic structures only, even though exhaustive argumentation against this prejudice has been given since a long time[18][19] (this is analogous to claiming that plane waves can be used for infinite uniform media only). Second, usually the Bloch expansion analysis of finite structures is made using a *global* plane wave expansion[14][15], that is a Fourier expansion on harmonics of the reciprocal lattice wavenumber, instead of a *local* plane wave expansion on the propagation constant of each layer. But, in the same way as multiplying *N* transfer matrices means forcing the local plane wave analysis at the global level (i.e. Transfer Matrix is blind to periodicity), using a global infinite (or, in practice, truncated) plane wave expansion means forcing the global periodicity at the local level (i.e. the basis elements of a Fourier expansion are blind to the period microstructure). We prefer instead to use the Bloch modes at the global level and to project them to local plane waves (i.e. local eigenmodes) whenever a local calculation is needed, without any truncation of any basis.

For a comparison with the quasinormal modes approach, we can find an analogy with the standard treatment of Fabry-Pérot cavities. A Fabry-Pérot can be seen both as an open cavity, i.e. a cavity with its proper "standing waves" (the quasinormal modes[20]) or, on the other hand, as a cascading of two mirrors that reflect progressive and regressive traveling waves, i.e. the eigenmodes of the medium that they surround. These two points of view apply also to periodic structures: the first one by finding the proper quasinormal modes[17], the second one by replacing the plane waves with the Bloch waves, that is what we have done.

As a numerical example we show in Fig.3 the transmission spectrum of a 10-period bilayer structure and its comparison with a standard transfer matrix approach. The complete agreement is not a surprise, beeing both exact methods, but we like to point out that our analytic approach is much more efficient because the calculation time does not depend on the number of periods. We also show the square modulus of the Bloch modes coefficients $X_\pm$: notice the asymptotes corresponding to the singularities at the band edges. In Fig. 4 are plotted the field amplitudes of both the Bloch wave decomposition (Eq. 6) and the local plane wave decomposition (Eq. 7) calculated at the first transmission peak on the right hand side of the second band (corresponding to $l = 1.5715 \mu m$). Notice the discrete invariance of the Bloch mode amplitudes from period to period and the continuous invariance of the plane wave amplitude within each single layer.
Two movies showing a comparison between the plane wave and the Bloch wave expansions are available on the web[21].

**B. Fractional period case: left to right**

It is also useful to calculate the field when the structure begins with the medium $j$ and finishes with the medium 1 (Fig. 5). Supposing to keep the same frame as in the previous case and defining the period fraction $x_j \equiv \Lambda/z_{j,0}$, if the incoming plane wave comes from the left hand side, the boundary conditions become:

$$\begin{cases} A_j^f(z_{j,0}) = 1 \\ A_j^b(z_{j,0}) \equiv \boldsymbol{r}^R(N-x_j) \\ A_1^f(N\Lambda) \equiv \boldsymbol{t}^R(N-x_j) \\ A_1^b(N\Lambda) = 0 \end{cases}$$

to give

$$\begin{cases} X_+ = \Omega_{1j}^N \exp(-iN\boldsymbol{q})h_1^- \\ X_- = -\Omega_{1j}^N \exp(iN\boldsymbol{q})h_1^+ \\ \boldsymbol{t}^R(N-x_j) = \Omega_{1j}^N(g_1^+ h_1^- - g_1^- h_1^+) \\ \boldsymbol{r}^R(N-x_j) = \Omega_{1j}^N[h_1^- h_j^+ \exp(-iN\boldsymbol{q}) - h_1^+ h_j^- \exp(iN\boldsymbol{q})] \end{cases} \quad (10)$$

where $\Omega_{1j}^N$ is defined in Eq. (9).

**C. Fractional period case: right to left**

When the structure begins with the medium 1 and finishes with the medium $j$ and it is excited from the right hand side (Fig. 6) the boundary conditions become:

$$\begin{cases} A_j^f(z_{j,N}) \equiv \mathbf{r}^L(N-x_j) \\ A_j^b(z_{j,N}) = 1 \\ A_1^f(N\Lambda) = 0 \\ A_1^b(N\Lambda) \equiv \mathbf{t}^L(N-x_j) \end{cases}$$

to give

$$\begin{cases} X_+ = -\Omega_{j1}^N g_1^- \\ X_- = \Omega_{j1}^N g_1^+ \\ \mathbf{t}^L(N-x_j) = \Omega_{j1}^N (g_1^+ h_1^- - g_1^- h_1^+) \\ \mathbf{r}^L(N-x_j) = \Omega_{j1}^N [g_1^+ g_j^- \exp(-iN\mathbf{q}) - g_1^- g_j^+ \exp(iN\mathbf{q})] \end{cases}, \quad (11)$$

where $\Omega_{j1}^N$ is defined in Eq. (9).

Notice the role exchange between $g_j^\pm$ and $h_j^\mp$ with respect to Eq. (10).

## IV. SECOND HARMONIC GENERATION

Having an analytical expression for the field allows to easily solve nonlinear problems in a finite periodic structure without using heavy numerical approaches. As an example we will show how to calculate analytically the second harmonic generated by a *N*-period multilayer in the undepleted pump regime.

## A. Nonlinear output

Suppose that to each layer of refractive index $n_j$ corresponds a nonlinear second order susceptibility $c_j^{(2)}$. The field at frequency ω and propagation constant $\mathbf{b}_j$ expressed by Eq. (7) will generate in each layer a time dependent material polarization[22]

$$\tilde{P}_j(z,t) \propto c_j^{(2)}[E_j(z)\exp(i\mathbf{w}t)+c.c.]^2.$$

We will keep only the two spatial terms which can significantly contribute to SHG, i.e.:

$$P_j^f(z) \propto c_j^{(2)}[A_j^f(z)]^2 \exp[-2i\mathbf{b}_j\mathbf{d}_j(z)]$$
$$P_j^b(z) \propto c_j^{(2)}[A_j^b(z)]^2 \exp[2i\mathbf{b}_j\mathbf{d}_j(z)]$$

So, at the right edge of the layer, the forward propagating field at frequency 2ω and propagation constant $k_j$, generated starting from the left edge will be (in each layer the period number $q(z) \equiv m$ and the field complex "amplitudes" $A_j^{f,b}(z) \equiv A_{j,m}^{f,b}$ are constant)

$$G_{j,m}^f \propto c_j^{(2)}(A_{j,m}^f)^2 \exp(-ik_j d_j/2) \operatorname{sinc}[(\mathbf{b}_j - k_j/2)d_j]\, d_j \qquad (12)$$

while, at the left edge of the layer, the backward propagating second harmonic field generated starting from the right edge will be

$$G_{j,m}^b \propto c_j^{(2)}(A_{j,m}^b)^2 \exp(-ik_j d_j/2) \operatorname{sinc}[(\mathbf{b}_j - k_j/2)d_j]\, d_j \qquad (13)$$

So, from the point of view of the 2ω radiation, each layer will be seen (see Fig. 7) as a cavity with two source terms at its edges, delimited by two mirrors with reflectivity $r_{j,m}^{R,L}$ and transmissivity $t_{j,m}^{R,L}$. For the layer $j$, from Eq. (10) with all quantities calculated at frequency $2\omega$, it will be

$$\begin{aligned}
t_{j,m}^{R} &= \exp(ik_j d_j/2) t^{R}(N-m-x_j) \\
r_{j,m}^{R} &= \exp(ik_j d_j/2) r^{R}(N-m-x_j) \\
t_{j,m}^{L} &= \exp(ik_j d_j/2) t^{L}(m+x_j) \\
r_{j,m}^{L} &= \exp(ik_j d_j/2) r^{L}(m+x_j)
\end{aligned},$$

where the phase factors account for propagation from the edges to the center of the layer. So it is possible to easily calculate the second harmonic contributions coming from the $j^{\text{th}}$ layer of the $m^{\text{th}}$ period by simply imposing the boundary conditions for any given cavity:

$$\begin{cases}
F_{j,m}^{f}(0) = r_{j,m}^{L} F_{j,m}^{b}(0) \\
F_{j,m}^{f}(d_j) = F_{j,m}^{f}(0) \exp(-ik_j d_j) + G_{j,m}^{f} \\
F_{j,m}^{b}(0) = F_{j,m}^{b}(d_j) \exp(-ik_j d_j) + G_{j,m}^{b} \\
F_{j,m}^{b}(d_j) = r_{j,m}^{R} F_{j,m}^{f}(d_j) \\
F_{j,m}^{R} = t_{j,m}^{R} F_{j,m}^{f}(d_j) \\
F_{j,m}^{L} = t_{j,m}^{L} F_{j,m}^{b}(0)
\end{cases},$$

where (see Fig. 7) $F_{j,m}^{f,b}$ are the forward and backward propagating fields inside the cavity and, solving for the fields $F_{j,m}^{R,L}$ out of the structure, we find:

$$\begin{cases}
F_{j,m}^{R} = t_{j,m}^{R} \dfrac{r_{j,m}^{L} \exp(-ik_j d_j) G_{j,m}^{b} + G_{j,m}^{f}}{1 - r_{j,m}^{L} r_{j,m}^{R} \exp(-2ik_j d_j)} \\
F_{j,m}^{L} = t_{j,m}^{L} \dfrac{r_{j,m}^{R} \exp(-ik_j d_j) G_{j,m}^{f} + G_{j,m}^{b}}{1 - r_{j,m}^{L} r_{j,m}^{R} \exp(-2ik_j d_j)}
\end{cases}.$$

Finally the total second harmonic field outside the multilayer will be simply given by the sum

$$F^{R,L} \equiv \sum_{j,m} F_{j,m}^{R,L},$$

which is the desired result (where, for simplicity, we have assumed also the thickness of the first and the last layer to be $d_1$).

**B. Field distribution inside the structure**

We can also derive an expression for the second harmonic field distribution inside the structure. Let's define the field generated by the $j^{\text{th}}$ layer of the $m^{\text{th}}$ period on the right hand side as

$$E_{j,m}^{R}(z) = X_{j,m}^{R,+} E^{+}(z) + X_{j,m}^{R,-} E^{-}(z) \quad z > z_{j,m},$$

and on the left hand side as

$$E_{j,m}^{L}(z) = X_{j,m}^{L,+} E^{+}(z) + X_{j,m}^{L,-} E^{-}(z) \quad z < z_{j,m}.$$

We can determine $X_{j,m}^{RL,\pm}$ by imposing the boundary conditions

$$\begin{cases} E_{j,m}^{R,f}(N\Lambda) = F_{j,m}^{R} \\ E_{j,m}^{R,b}(N\Lambda) = 0 \end{cases}, \quad \begin{cases} E_{j,m}^{L,f}(0) = 0 \\ E_{j,m}^{L,b}(0) = F_{j,m}^{L} \end{cases},$$

having defined

$$E_{j,m}^{R,f}(z) \equiv X_{j,m}^{R,+} g_1^+ \exp[iqq(z)] + X_{j,m}^{R,-} g_1^- \exp[-iqq(z)]$$

$$E_{j,m}^{R,b}(z) \equiv X_{j,m}^{R,+} h_1^+ \exp[iqq(z)] + X_{j,m}^{R,-} h_1^- \exp[-iqq(z)]$$

$$E_{j,m}^{L,f}(z) \equiv X_{j,m}^{L,+} g_1^+ \exp[iqq(z)] + X_{j,m}^{L,-} g_1^- \exp[-iqq(z)]$$

$$E_{j,m}^{L,b}(z) \equiv X_{j,m}^{L,+} h_1^+ \exp[iqq(z)] + X_{j,m}^{L,-} h_1^- \exp[-iqq(z)]$$

to give

$$\begin{cases} X_{j,m}^{R,+} = \Omega_{11}^0 h_1^- \exp(-iN\mathbf{q}) F_{j,m}^R \\ X_{j,m}^{R,-} = -\Omega_{11}^0 h_1^+ \exp(iN\mathbf{q}) F_{j,m}^R \end{cases}, \quad \begin{cases} X_{j,m}^{L,+} = -\Omega_{11}^0 g_1^- F_{j,m}^L \\ X_{j,m}^{L,-} = \Omega_{11}^0 g_1^+ F_{j,m}^L \end{cases},$$

where $\Omega_{11}^0$ is defined in Eq. (9), always with all quantities calculated at frequency $2\omega$.

Since in the $i^{\text{th}}$ layer of the $n^{\text{th}}$ period will contribute all the $X_{j,m}^{R,\pm}$ so that $m < n$ OR ($m = n$ AND $j \leq i$) as well as all the $X_{j,m}^{L,\pm}$ so that $m > n$ OR ($m = n$ AND $j \geq i$), the overall Bloch mode coefficients will depend on the position and will be:

$$X_\pm^{i,n} \equiv \sum_{m<n} \sum_j X_{j,m}^{R,\pm} + \sum_{j \leq i} X_{j,n}^{R,\pm} + \sum_{m>n} \sum_j X_{j,m}^{L,\pm} + \sum_{j \geq i} X_{j,n}^{L,\pm}$$

so that we can write the total second harmonic field in the layer as:

$$E_{i,n}(z) = X_+^{i,n} E^+(z) + X_-^{i,n} E^-(z)$$

which is the desired result.

## V. CONCLUSIONS

In the contest of the Translation Matrix formalism, we have shown how to obtain analytic expression for the field in finite 1D periodic structures. We apply this result to the case of SHG which can be treated analytically in the undepleted pump regime. The same analysis can be extended to much more complicated structures (e.g. featuring defects) by simply imposing the boundary conditions that apply or by associating the proper transfer matrix to each selfsimilar section (i.e. regarding each periodic section as a concentrated mirror). Also other optical harmonic generations can be treated in the undepleted pump approximation, and pump depletion could be taken into account with a numerical iteration scheme[23]. We expect that this approach, which relies only on 2-dimensional bases, can be implemented to give very efficient numerical tools for the study of multilayer structures and related phenomena.

## ACKNOWLEDGEMENTS

I thank D. Faccio, M. Tormen for useful discussions and for encouraging this work. I thank also M. Centini for fruitful discussions about Bloch modes and quasinormal mode expansion.

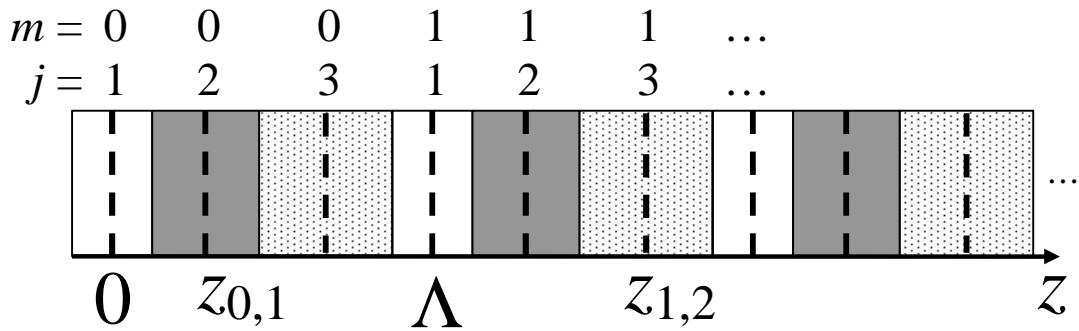

**FIGURE 1. A periodic multilayer, which period is composed by three layers. On top is explained the meaning of the indexes *m*, indicating the period, and *j*, indicating the layer.**

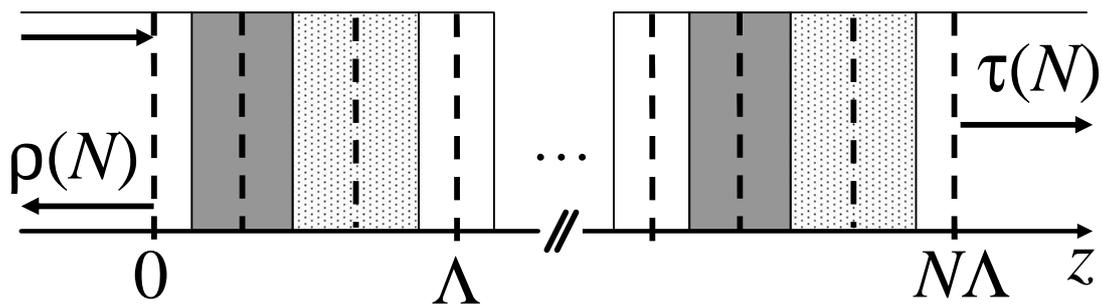

**FIGURE 2. Incidence on a *N*-period structure. The light impinging from the left is partially transmitted and partially reflected with amplitude and phase given by the complex numbers t and r.**

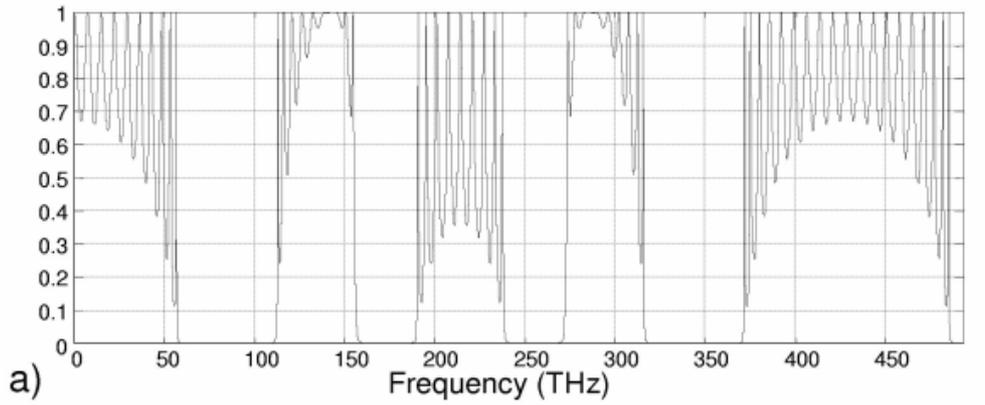

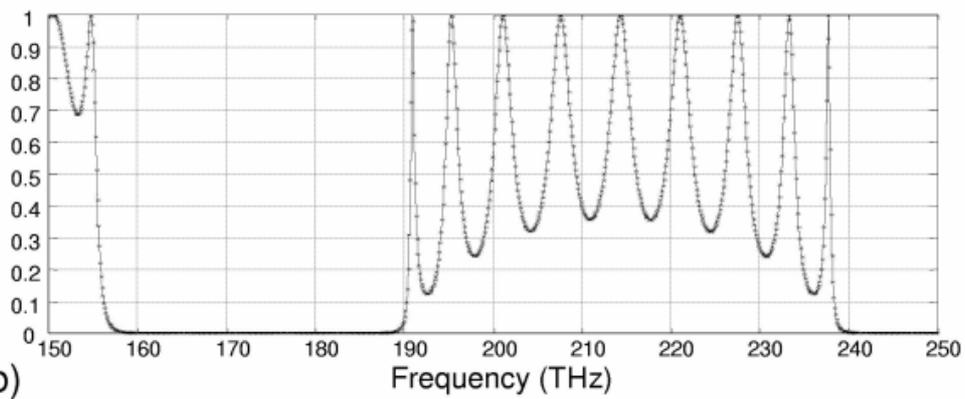

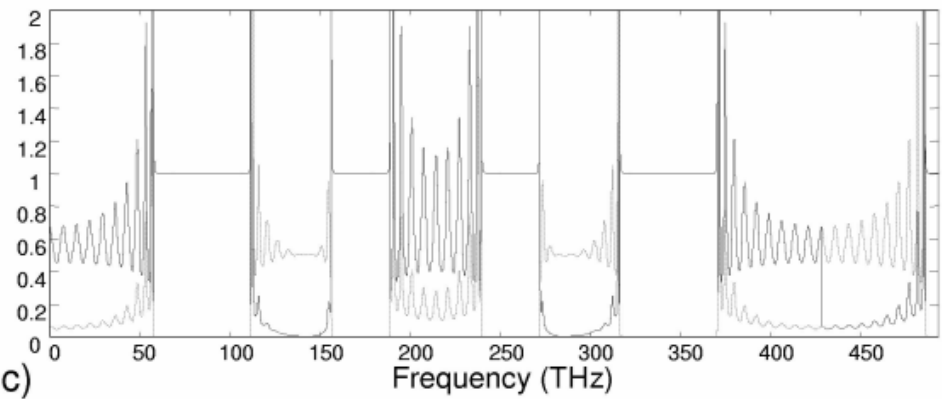

**FIGURE 3.** a) Transmission spectrum of a 10-period structure ($d_1$=700nm, $d_2$=350nm, $n_1$=1, $n_2$=3) and b) its comparison with a standard transfer matrix approach (crosses); c) represents the square modulus of the Bloch modes coefficients $X_\pm$.

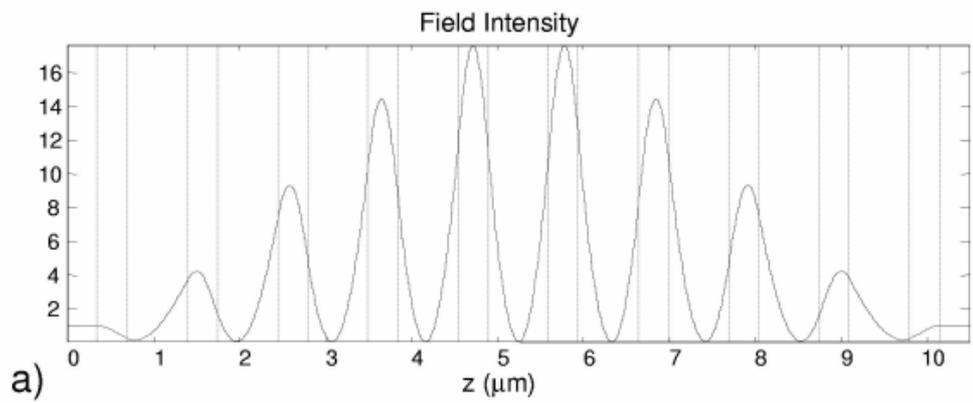

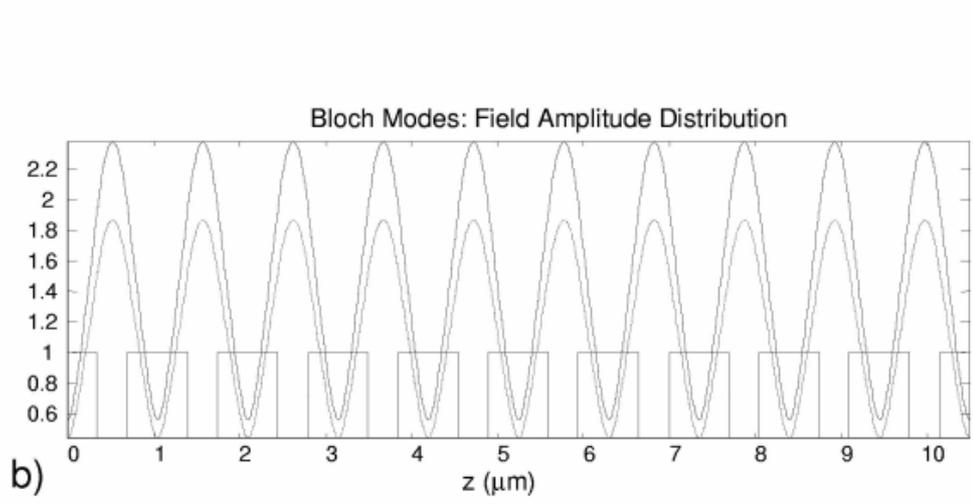

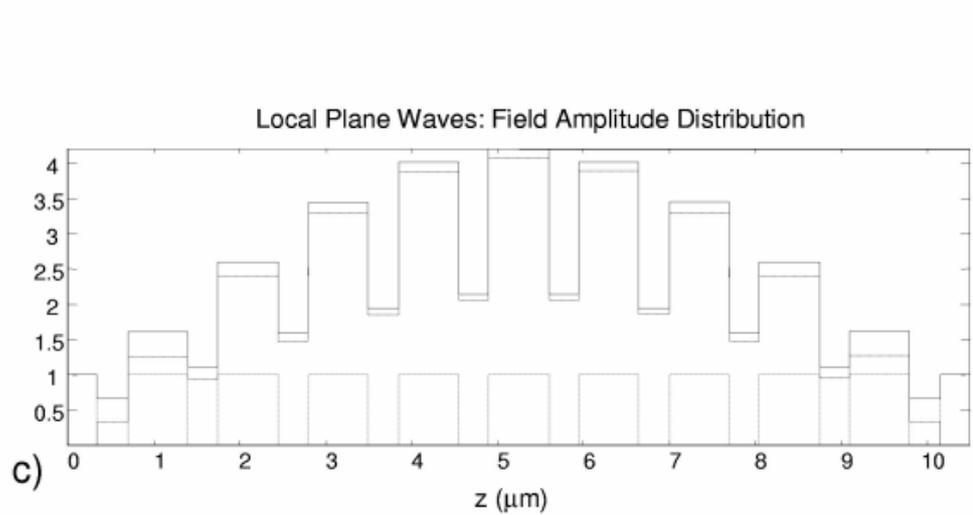

**FIGURE 4. a) Total electric field intensity, b) field amplitudes of the Bloch wave decomposition and c) Field amplitudes of the local plane wave decomposition, all calculated at the first transmission peak on the right hand side of the second band as in Fig. 3.**

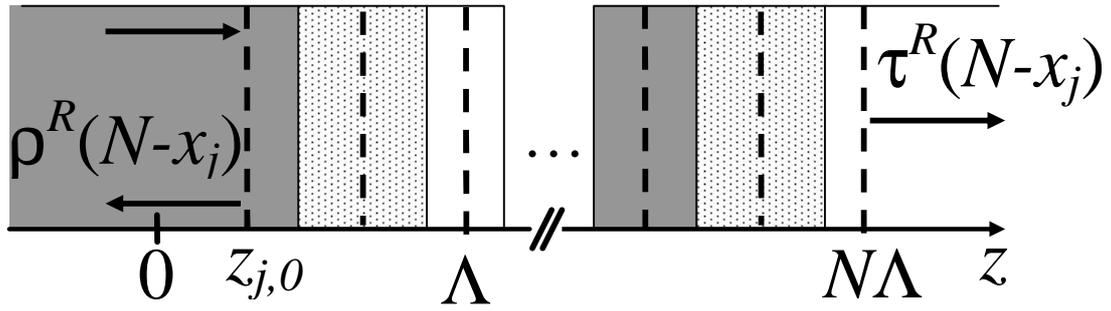

**FIGURE 5.** Left to right incidence on a $(N-x_j)$-period structure. The first period is truncated and begins with the $j^{th}$ layer.

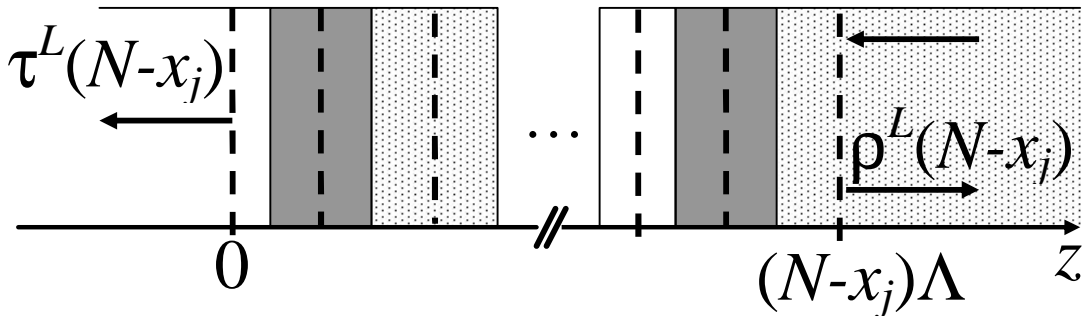

**FIGURE 6.** Right to left incidence on a $(N-x_j)$-period structure. The last period is truncated and finishes with the $j^{th}$ layer.

**FIGURE 7.** Equivalent cavity for the light generated in a single layer. The generation can be thought to be concentrated in the point sources *G* and to build up the fields *F* inside and outside a cavity. The concentrated mirrors are equivalent to the multilayers on the left and on the right of the given layer.

[1] P. Yeh, A. Yariv, and C. Hong, J. Opt. Soc. Am. **67**, 438 (1977).

[2] P. Yeh, *Optical Waves in Layered Media* (Wiley, New York, 1988).

[3] M. A. Muriel and A. Carballar, IEEE Photon. Technol. Lett. **7**, 955 (1997).

[4] D. S. Bethune, J. Opt. Soc. Am. B **6**, 910 (1989).

[5] S. Enoch and H. Akhouayri, J. Opt. Soc. Am. B **15**, 1030 (1998).

[6] M. Midrio, L. Socci, and M. Romagnoli, J. Opt. Soc. Am. B **19**, 83 (2002).

[7] H. Kogelnik, Bell Syst. Tech. J. **48**, 2909 (1969).

[8] M. G. Moharam and T. K. Gaylord, J. Opt Soc. Am. **71**, 811 (1981).

[9] Z. Zylberberg and E. Marom, J. Opt. Soc. Am. **73**, 392 (1983).

[10] C. M. de Sterke and J. E. Sipe, Phys. Rev. A **38**, 5149 (1988).

[11] J. E. Sipe, I. Poladian, and C. Martijn de Sterke, J. Opt. Soc. Am. A **11**, 1307 (1994).

[12] J. W. Haus, R. Viswanathan, M. Scalora, A. Kalocsai, J. D. Cole, and J. Theimer, Phys. Rev. A **57**, 2120 (1998).

[13] G. D'Aguanno, M. Centini, M. Scalora, C. Sibilia, M. Bertolotti, M. J. Bloemer, and C. M. Bowden, J. Opt. Soc. Am. B **19**, 2111 (2002).

[14] P. St. J. Russell, T. A. Birks, and F. D. Lloyd-Lucas, "Photonic Bloch Waves and Photonic Band Gaps", in *Confined Electrons and Photons*, edited by E. Burstein and C. Weisbuch, (Plenum, New York, 1995), pp 585.

[15] P. St. J. Russell, Appl. Phys. B **39**, 231 (1986).

[16] D. Faccio, F. Bragheri, and M. Cherchi, J. Opt. Soc. Am. B (to be published).

[17] A. Settimi, S. Severini, N. Mattiucci, C. Sibilia, M. Centini, G. D'Aguanno, M. Bertolotti, M. Scalora, M. Bloemer, and C. M. Bowden, Phys. Rev. E **68**, 026614-1 (2003).

[18] P. St. J. Russell, J. Opt. Soc. Am. A **1**, 293 (1984).

[19] P. St. J. Russell, Opt. Comm. **48**, 71 (1983).

[20] P. T. Leung, S. Y. Liu, and K. Young, Phys. Rev. A **49**, 3057 (1994).

[21] EPAPS: Bloch.avi, Plane.avi

[22] R. W. Boyd, *Nonlinear Optics* (Academic, San Diego, 1992).

[23] Y. Jeong and B. Lee, IEEE J. Quantum Electron. **35**, 162 (1999).